\documentclass[referee]{aa}

\usepackage{graphicx}

\usepackage{natbib}

\bibpunct{(}{)}{;}{a}{}{,}
\usepackage{amsmath}
\usepackage{txfonts}

\begin{document}


\title{Magnetic clouds in the solar wind: A numerical assessment study
  of analytical models}

\author{Giorgi Dalakishvili \inst{1,2,3}
  \and Jens Kleimann \inst{1}
  \and Horst Fichtner \inst{1}
  \and Stefaan Poedts \inst{2}
}

\institute{
  Institut f\"ur Weltraum- und Astrophysik, Ruhr-Universit\"at Bochum, Germany\\
  \email{giorgi@tp4.rub.de}, \email{jk@tp4.rub.de}, \email{hf@tp4.rub.de}
  \and
  Centre for Plasma Astrophyics, Katholieke Universiteit Leuven, Belgium\\
  \email{Stefaan.Poedts@wis.kuleuven.be}
  \and
  Center for Theoretical Astrophysics, Institute of Theoretical Physics,
  Ilia State University, Tbilisi, Georgia\\
}

\abstract
{Magnetic clouds (MCs) are ``magnetized plasma clouds" moving in the solar
  wind. MCs transport magnetic flux and helicity away from the Sun. These
  structures are not stationary but feature temporal evolution as they
  propagate in the solar wind. Simplified analytical models are frequently
  used for the description of MCs, and fit certain observational data well.
}
{ \textbf{
    The goal of the present study is to investigate numerically the validity of
    an analytical model which is widely used for the description of MCs, and
    to determine under which conditions this model's implied assumptions cease
    to be valid.
  } }
{A numerical approach is applied. Analytical solutions that have been derived
  in previous studies are implemented in a \textbf{3-D magnetohydrodynamic}
  simulation code as initial conditions.}
{Initially, the analytical model represents the main
  observational features of the MCs. However, these characteristics
  prevail only if the structure moves with a velocity close to
  the velocity of the background flow. In this case an MC's evolution can
  quite accurately be described using an analytic, self-similar approach.
  The dynamics of the magnetic structures which move with a velocity
  significantly above or below that of the velocity of the solar wind
  is investigated in detail. Besides the standard case in which MCs only
  expand and propagate in the solar wind, the case of an MC rotating around
  its axis of symmetry is also considered, and the resulting influence
  on the MC's dynamics is studied.}
{A comparison of the numerical results with observational data indicates
  reasonable agreement especially for the intermediate case, in which the
  MC's bulk velocity and the velocity of the background flow are equal.
  In this particular case, analytical solutions obtained on the basis of a
  self-similar approach indeed describe the MC's evolution quite accurately.
  In general, however, numerical simulations are necessary to investigate
  the evolution as a function of a wide range of the parameters which
  define the initial conditions.}


\titlerunning{Magnetic clouds in the solar wind}

\authorrunning{Dalakishvili et al.}
\maketitle

\section{Introduction}

It is well-known that \textit{coronal mass ejections} (CMEs) are one
of the most significant forms of solar activity. They carry enormous
masses of plasma, threaded by a magnetic field, away into the
interplanetary medium. Further away from the Sun, these large-scale,
dynamical plasma structures are commonly called
\textit{interplanetary coronal mass ejections} (ICMEs). A magnetic
cloud (MC) is a specific type of ICME
\citep[see e.g.][]{bur95,wim06,dem08} and can be considered as a
magnetically isolated structure moving in the solar wind. Three
features of such magnetic structures -- an enhanced magnetic field,
the rotation of this magnetic field and the low proton temperatures
-- are selected as \textit{bona fide} signatures of MCs
\citep{bur95}. \textit{In situ} observations of these physical
properties of MCs are considered as important prerequisites towards
a prediction of the geophysical effectiveness of their interaction
with the Earth's magnetosphere, i.e. for space weather forecasts and
related issues.

Different models for the structures of MCs have been proposed. While
there is no general agreement about their large-scale structure, the
local structure of MCs is commonly considered in the form of
cylindrically symmetric force-free configurations
\citep{bur88, bur91, dem09}. It is
often suggested that the ends of MCs connect to the solar surface
while, according to some other models, MCs are described as tori
\citep{RVP06,RVP07}. These models can be useful for capturing
particular features of MCs, such as the curvature of an MC's axis.
In a number of studies, MCs are considered as force-free, static,
axially symmetric rigid flux ropes and their magnetic field is
constructed on the basis of Lundquist's model
\citep{bur88, lep90, far93}. None of
these analytical studies do consider interactions of MCs with the
ambient environment. Observations show, however, that MCs do not
stay static but expand while propagating in the solar wind, they
keep expanding well beyond 1~AU, and they dynamically interact with
the background solar wind flow
\citep{bur91, bot98, dem08, dem09}.

Analytical models describing the features of MCs
are used with remarkable success by a number of experts in order to
describe certain observational data. Derivations of the analytical
expressions for physical variables characterizing MCs are based on
the assumption that a MC represents a self-similarly evolving
cylindrical structure
\citep{dal10, dem09, nak08, far95, far93, lep90, bur88}. We find it
appropriate and necessary to numerically investigate the validity of
such analytical models. For this purpose, it is seems reasonable to
start out with a comparably simple model based on analytical
considerations, and then to successively relax these simplifications
towards more realistic configurations. This procedure is then able
to reveal the extent to which simplifications such as the assumption
of self-similarity maintain their validity in more complex settings.

Several numerical studies
\citep{desterck1, desterck2, man04, chane06, dal09}
have been performed to
explore character of magnetized plasma flows near MCs by treating them as
superconductors, i.e.\ the magnetic field does not penetrate the
cylindrical structure. \citet{ow05} have revealed that the region in front
of an interplanetary MC has a rather complicated
structure. In order to gain more insight into the characteristics of
this region and the physical phenomena within it, and in order to
better understand the geo-effectiveness of MCs, a refined
investigation of MC dynamics is required.

For instance, it is to be expected (and has in fact been
demonstrated numerically by, e.g.\ \citet{Ods99} that
variations in the structure of the background wind can severely
distort the initially simple geometry of a MC. Furthermore, \citet{ril04}
have used MHD simulations as well as kinematic
arguments to show that MCs tend to flatten out as they propagate
outwards, but still stress the paramount significance of force-free
field models for the interpretation of MC observations. It is thus
of vital importance to establish under which conditions (if any)
these simple force-free models continue to be applicable. The need
to eliminate all secondary effects to access the true net effect of
a perturbing magnetic cloud on the ambient solar wind justifies our
deliberate choice of both a cylindrical MC geometry and an
unstructured background flow (see Section~3) as a first step towards
this goal.

The existence of analytical models properly describing the evolution
of MCs is very valuable for the field. \citet{dal10} numerically
investigated the evolution of
self-similar analytical solutions and showed that the solutions
maintain a self-similar structure for a comparatively long time.
However, in this 1-D study the entire structure was considered to be
cylindrically symmetric, and it was assumed that the background
solar wind could be described on the basis of a self-similar
approach. In the present study we employ a 3-D code and implement a
more general background flow.

In recent years, various numerical studies have been performed to
investigate the initiation and the dynamics of CMEs
\citep{bart05, bart07, jacobs07, jens05, jens09, asch08, ril08,
amar10}. These authors studied the initiation and propagation of
solar eruptions in the heliosphere, and followed the evolution of
magnetic structures in a 3-D setting. The numerical solutions show
that far from the Sun, the steady-state solar wind can be
characterized by a radial velocity and, to some approximation, by a
radial magnetic field. Therefore, in our simulations we employed a
radial flow with a radial magnetic field as an initial background in
order to study the evolution of MCs at larger heliocentric
distances. The diameter of the MC is only 0.2~AU at 1~AU, so we
could assume the physical characteristics of the ambient solar wind
not to change significantly in the region where it interacts with
the MC.


Additionally, since a rotation of MCs about its axis of symmetry is
deduced from observations
\citep{bur95, farb95, kl82}, we found it \textit{necessary to start
an investigation of the dynamics of MCs taking into account this
rotation. This is possible by analytically formulating an
appropriate initial set-up.}
 As stated above, many groups have
simulated MCs, and various tests and comparisons of numerical models
with each other and with measurements have been performed, see, e.g.
\citet{van10, van09}
and reference therein. It appears, however,
that comprehensive, systematic comparisons of analytical models of
MCs with their full three-dimensional numerical simulations have not
been made -- there are only a few studies into this direction. For
example, \citet{van00} have compared the analytical
model by \citet{osher95} with two-dimensional numerical
simulations and found very good agreement. Although the analytical
model served mainly as a test case for the numerical simulations,
\citet{van00} 
came to the conclusion that the
analytically determined asymptotic behavior of the magnetic field
on the axis of an expanding flux tube has a rather general validity.

 Other authors have compared numerical and analytical results of
certain aspects of MC physics, see e.g.\ \citet{xiong06}, who
studied the geo-effectiveness of so-called shock-overtaking MCs with
a 2.5-dimensional model. More recently, \citet{taubenschuss} have
also performed 2.5-dimensional simulations and compared them
regarding the expansion speed to some extent with analytical
findings by \citet{ow05}, resulting in average agreement.
Interestingly, these authors also claim that their simulations
reveal that the force-free configuration for MCs seems to be
conserved very well, at least when averaging over the entire cross
section.

With the present analysis, we compare a recently developed
analytical model with three-dimensional numerical simulations.
The remainder of this paper is organized as follows: in Section~2 we
describe the analytical and in Section~3 the numerical model,
including the implemented initial and boundary conditions. In
Section~4 we describe details of the simulation setup. In Section~5,
the results of our simulations are presented and discussed,
and in Section~6 we summarize the results of our work and discuss
the conclusions as well as indicate future plans.

\section{Analytical background}

In this section we briefly summarize the analytical model developed
in \citet{dal10}. To perform an analytic study of the dynamics of
magnetic clouds, we have to start from the full set of ideal MHD equations

\begin{align}
  \label{eq:1}
  \nabla \cdot \vec{B} &= 0 , \\
  \partial_t \vec{B} &= \nabla \times \left[ \vec{V} \times \vec{B} \right] \\
  \partial_t \varrho + \nabla \cdot \left( \varrho \vec{V} \right) &= 0 \\
  \label{eq:4}
  \varrho \left[ \partial_t +
    \left( \vec{V} \cdot \nabla \right) \right] \vec{V}
  &= \mu^{-1} (\nabla \times \vec{B}) \times \vec{B} -\nabla p
\end{align}

In these equations, $p$ denotes the thermal plasma pressure, $\varrho$
is the mass density, $\vec{V}$ is the velocity field, $\vec{B}$ denotes
the magnetic field, and $\mu$ is the permeability of free space.\\
The problem is considered in the frame of the MC and in
cylindrical coordinates centered on the MC, i.e.\ with a
longitudinal axis $z$ that coincides with the MC's axis. In a number
of previous studies, the MCs were considered as cylindrical magnetic
structures, characterized by axial symmetry. In the present
consideration, both symmetry along the $z$ axis ($\partial_z=0$)
and the azimuthal direction ($\partial_{\varphi}=0$) are assumed.
The axially symmetric magnetic field can then be expressed as

\begin{equation}
\vec{B} \equiv [\ 0, \ B_{\varphi}, \ B_{z}]
\end{equation}

where $B_{\varphi} = B_{\varphi}(r,t)$ and $B_{z}=B_{z}(r,t)$. We
note
that this representation satisfies the solenoidal condition (\ref{eq:1}).\\
The self-similar approach, adopted here, implies that the temporal
evolution of the physical functions is controlled by the
self-similarity variable

\begin{equation}
\xi \equiv \frac{r}{\Phi(t)}
\end{equation}

where $\Phi(t)$ is a function of time. In analogy to \citet{low82},
let us search solutions of the MHD equations in the following form:

\begin{align}
  \label{eq:7a}
  B_{\varphi} &= \Phi^{\delta} Q_{\varphi}(\xi ) \tag{7a} \\
  B_{z}      &= \Phi^{\sigma} Q_{z}(\xi)        \tag{7b} \\
  \varrho    &= \Phi^{\alpha} \tilde \rho(\xi) \tag{7c} \\
  \label{eq:7d}
  p          &= \Phi^{\beta} \tilde p(\xi)     \tag{7d}
\end{align}

\setcounter{equation}{7}

One can see that the type of solutions introduced by
Eqs.~(\ref{eq:7a}-\ref{eq:7d}) evolve self-similarly and are
characterized by a particular time-scaling. Here $Q_{\varphi}$,
$Q_{z}$, $\tilde \rho$, and $\tilde p$ are functions of the
self-similar variable $\xi$, and $\Phi^{\delta}$, $\Phi^{\sigma}$,
$\Phi^{\alpha}$, and $\Phi^{\beta}$ show the time scaling of the
azimuthal and longitudinal components of the magnetic field, the
plasma density,
and the plasma pressure, respectively. \\
We consider both a radial and a longitudinal expansion of the MC, but
no motion in the azimuthal direction. In this case, the
Eulerian velocity field of the plasma, $\vec{V}$, can be expressed as

\begin{equation}
  \vec{V} = \left[ V_{r}, \ 0, \ V_{z} \right] ,
\end{equation}

where we assume that the radial component of the velocity
$V_{r}=V_{r}(r,t)$, and the $z$ component $V_{z}=V_{z}(z,t)$, i.e.\
we assume that the MC maintains its cylindrical shape during its
evolution. \\
The solutions readily follow from the derived equations
\citep[see][]{dal10}, yielding

\begin{align}
  \label{eq:9a}
  \vec{V}_{r,\varphi,z} &= \frac{k}{1+k t} \left[ r,0,z \right] \tag{9a} \\
  \varrho   &= \frac{\tilde \varrho}{(1+k t)^{3}}  \tag{9b} \\
  p & = \frac{\tilde p}{(1+k t)^4} \tag{9c}
\end{align}

where $\tilde \varrho$ and $\tilde p$ are arbitrary functions of
$\xi=r/\Phi$. The components of the magnetic field read

\setcounter{equation}{9}

\begin{equation}
\label{eq:10}
  \vec{B}_{r\varphi z} = \frac{B_{0}}{(1+k t)^{2}} \left[ \ 0,
    \ J_{1} \left(\frac{r}{\tilde r_{0}(1+k t)}\right),
    \ J_{0} \left(\frac{r}{\tilde r_{0}(1+k t)}\right) \right] ,
\end{equation}

where $J_{0}(x)$ and $J_{1}(x)$  are the Bessel functions of the
first kind while $B_{0}$ and $\tilde r_{0}$ are constants. Here $k$
is a constant parameter and $1/k$ has unit of time and characterizes
the MC's rate of radial expansion.

\section{Description of the model}

\subsection{Coordinate systems}
\begin{figure*}[t]
\vspace*{1mm}
\begin{center}
\includegraphics[width=15cm]{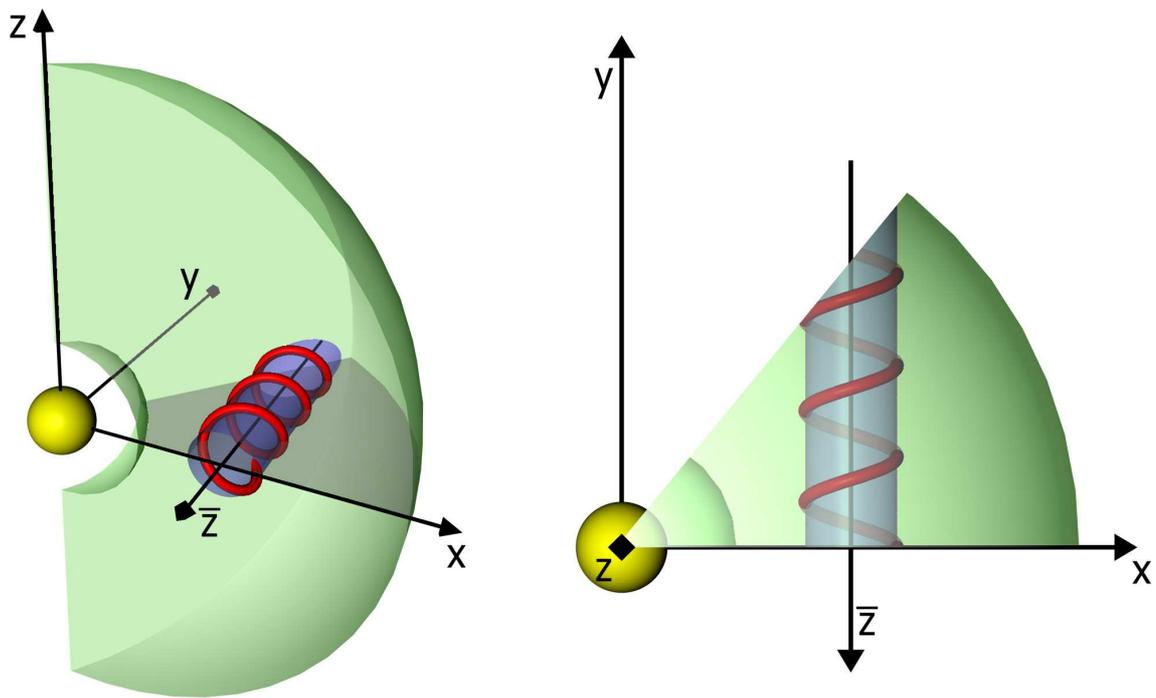}
\end{center}
\caption{ \label{fig:11}
  Sketch of the computational volume with the global,
  Sun-centered $(x,y,z)$ system, and the MC (cylinder with $\bar{z}$
  axis and exemplary field lines) in the ecliptic plane.}
\end{figure*}

It is assumed that an MC is initially a cylindrical structure placed
in the radial solar wind flow. The MC's initial bulk velocity is
perpendicular to its axis of symmetry. Hereafter, in order to
formulate the initial conditions with particular expressions in a
compact way, we introduce local and global coordinate systems. The
global coordinate system which we use is a spherical one
$(R,\vartheta,\varphi)$ centered on the Sun. The polar axis of this
system coincides with the solar magnetic axis (the $z$ axis), and
the azimuthal angle $\varphi$ is counted from the $x$ axis, which is
directed from the Sun to the Earth. The local, cylindrical
coordinate system ($r,\tilde{\varphi},\tilde{z}$) is related to a
cylindrical magnetic cloud: the $\tilde{z}$ axis coincides with the
axis of the MC and is perpendicular to the $(x,z)$ plane, such that
$(x,z,\tilde{z})$ defines a right-handed Cartesian coordinate
system. The azimuthal angle of the local system $\tilde{\varphi}$ is
counted from the $z$ axis of the global system. Fig.~\ref{fig:11}
shows a sketch of the computational volume and the respective
coordinate axes. After fixing the location of the cylinder in the
global system, we can define functional relations between the
coordinates of these two systems and transform vector components
from one system to another. Later, the notation
$C_{r,\tilde{\varphi},\tilde{z}}$ will indicate components of a
vector $\vec{C}$ in the local coordinate system, while
$C_{R,\vartheta,\varphi}$ will denote the components of the same
vector in the global coordinate system.

\subsection{Initial and boundary conditions}
The physical quantities are normalized as follows. The unit length
$L_{0}=7\cdot 10^{5}$~km is equal to the Solar radius. The unit magnetic
field $B_{0}=3$~nT, unit number density \mbox{$n_{0}=10$~cm$^{-3}$}, and
mass density \mbox{$\rho_{0}=m_{\rm p}n_{0}=1.7\cdot 10^{-20}$~kg/m$^{3}$}
(where $m_{\rm p}$ is the proton mass) approximate their respective values
at 1~AU. The speed is then normalized to
\mbox{$V_{\rm 0A}=B_{0}/\sqrt{\mu\rho_{0}}=20.5$~km/s} (i.e.\ of the
order of the MC's expansion velocity in the local frame, according
to \citet{van09}) and, finally, the unit of time is then given by
\mbox{$t_{0}=L_{0}/V_{0A}=9.5$~h}.
As a background plasma flow at large heliocentric distances we
consider a radial flow with a radial magnetic field. In order to
ensure the stationarity of this background flow, we choose the
following expressions for the initial background density and radial
magnetic field:

\begin{align}
  \label{eq:11a}
  \varrho_{\rm out} &= \varrho_{0, {\rm out}}
  \left( \frac{1}{R} \right)^{2} \tag{11a} \\
  \label{eq:11b}
  B_{{\rm out}, R, \vartheta, \varphi} &= B_{0, {\rm out}}
  \left(\frac{1}{R} \right)^{2}
  \left[ \Theta(\vartheta) , 0, \ 0 \right] . \tag{11b}
\end{align}

\setcounter{equation}{11} Here $\Theta(\vartheta)$ is an arbitrary

function of the polar angle. In our simulations, we considered both
the case $\Theta(\vartheta)=1$ and
$\Theta(\vartheta)=\cos\vartheta$. The latter case describes the
change of sign of the magnetic field and the existence of a current
sheet in the equatorial plane. We found that such asymmetry in the
magnetic field does not have a significant influence on the dynamics
of the MC.
Inside the MC, the force-free magnetic fields of Eq.~(\ref{eq:10}),
evaluated at $t=0$, is superimposed on background radial field:

\begin{equation}
  \vec{B}_{\rm in} = \vec{B}_{\rm in}^{\prime} + \vec{B}_{\rm out} \ .
\end{equation}

In the local coordinate system, the former becomes

\begin{equation}
  \label{eq:13}
  \vec{B}_{{\rm in} \ r, \tilde{\varphi}, \tilde{z}}^{\prime}
  = B_{0,{\rm in}} \left[0,
    \ J_{1}\left(\frac{r}{\tilde{r_{0}}} \right),
    \ J_{0}\left(\frac{r}{\tilde{r_{0}}} \right) \right] \ .
\end{equation}

We note that Eq.~(\ref{eq:13}) coincides with Lundquist's solution
\citep{lun50}.\\
Here, $\tilde{r_{0}}$ is a same
arbitrary constant parameter introduced in Eq.~(\ref{eq:10}). In the
numerical simulations, $\tilde{r_{0}}$ is equal to the MC's initial radius.
For a determination of the constants $B_{0,{\rm in}}$ and $B_{0 {\rm out}}$,
the observational data, which show that the plasma beta inside a MC
is lower than outside ($\beta_{\rm in} \ll 1$, $\beta_{\rm out}\sim 1$)
\citep{bur81, bur91, bot98} were taken into consideration.

The evolution of an MC is characterized by an increase of its radial
and longitudinal extent, and by the translational and rotational
motion of the whole structure in the ambient environment. The
velocity of matter inside the MC consists of the MC's bulk velocity
(which is initially perpendicular to the MC's axis of symmetry), the
velocity caused by an increase of its radius, and the velocity of
lengthening. Substituting $t=0$ in Eq.~(\ref{eq:9a}), we derive the
expressions for the components of the velocity inside the MC

\begin{equation}
  \label{eq:14a}
  \vec{V}_{{\rm in} \ r, \tilde{\varphi}, \tilde{z}}^{\prime}
  = k \left[r, \ 0, \ \tilde{z}\right]. \tag{14a}
\end{equation}

given in the coordinate system attached to the MC. The total
velocity of matter inside the MC is then

\begin{equation}
  \vec{V}_{\rm in} = \vec{V}_{\rm MC} + \vec{V}_{\rm in}^{\prime} \ . \tag{14b}
\end{equation}

The initial background velocity is prescribed to be radial and
constant far away from the MC, and tangential to the MC's surface,
i.e.\ the background plasma does not penetrate the MC. An expression
for the velocity which satisfies the above-mentioned conditions is:

\begin{equation}
  \vec{V}_{\rm out} = \vec{V} + \vec{V}^{\prime} \ . \tag{15a}
\end{equation}

where

\begin{equation}
  \vec{V} = (V_{R},0,0) \tag{15b}
\end{equation}

in the global coordinate system. It is convenient to express
$\vec{V}^{\prime}$ in local cylindrical coordinates:

\begin{align}
  V^{\prime}_{r} &= \left( V_{S}+V_{{\rm MC}, r} - V_{r} \right)
  \left(\frac{r_{0}}{r}\right)^{3} \tag{16a} \\
  \label{eq:16b}
  V^{\prime}_{\tilde{\varphi}} &= \left( V_{{\rm MC}, \tilde{\varphi}}
    - V_{\tilde{\varphi}}\right) \left(\frac{r_{0}}{r}\right)^{3} \tag{16b} \\
  V^{\prime}_{\tilde{z}} &= V_{{\rm MC}, \tilde{z}}
  \label{eq:16c}
  \left(\frac{r_{0}}{r}\right)^{3} \ . \tag{16c}
\end{align}

Here $V_{S}=k r_{0}$ is the initial Lagrangian velocity of the MC's
edge. The parameter $k$ is the constant introduced in
Eq.~(\ref{eq:9a}), see also \citet{dal10}. Initially, the density
inside the MC is uniform and half the unit density. The initial
radius of the MC is 20 in normalized units, which corresponds to a
value of 0.1~AU, which is confirmed to be reasonable by in-situ
measurements \citep{bur95}.\\
In addition to the kinematic case described by
Eqs.~(\ref{eq:14a}-\ref{eq:16c}), we also consider the interesting case in
which the MC rotates around its axis. In this case we change
Eq.~(\ref{eq:14a}) and Eq.~(\ref{eq:16b}) as follows:

\begin{align}
V^{\prime}_{{\rm in}, \tilde{\varphi}} &= \omega r \tag{17a} \\
V^{\prime}_{\tilde{\varphi}} &= \left(\tilde{V}_{{\rm MC},\tilde{\varphi}}
  -\tilde{V}_{\tilde{\varphi}} + \omega r \right)
\left(\frac{r_{0}}{r}\right)^{3} \ . \tag{17b}
\end{align}

Here $\omega$ is the angular velocity of the MC's rigid rotation
around its symmetry axis.


We applied the following boundary conditions formulated in the
global coordinate system: on the inner radial boundary we prescribe
for the magnetic field and density the functions given by
Eqs.~(\ref{eq:11a}-\ref{eq:11b}) and a radially uniform velocity. On
the other (opposite and other) boundaries, the density, the tangential
components of the magnetic field, and the velocity are extrapolated,
while for the normal components of the velocity and magnetic field
we use mass and magnetic flux conservation conditions. One should
bear in mind that the outer radial boundary conditions (whose choice was
motivated by simplicity) are incompatible with the cylindrical geometry
prescribed to the MC. While in principle one could expect an
influence of these boundary conditions on the solutions, they turn
out to be negligible far from the boundary: the propagation speed of
the magnetic structures is the local Alfv\'en speed.
Therefore, we could conclude that the boundary-induced disturbances
will not be able to propagate inwards against the supersonic outflow.

\subsection{Model equations and their implementation in the code}
In order to study the dynamics of MCs numerically, we use a
second-order finite volume scheme based on the work by \citet{kurg01}
for the hyperbolic part of the system of equations
\mbox{(\ref{eq:1}-\ref{eq:4})}, see also \citet{flaig09} and references
therein. This is a central conservative scheme for the solution of
equations of type
$\partial_t \vec{u} + \nabla \cdot [\vec{F(u)}] = \vec{0}$.
In our study we solve the following (normalized) equations:

\begin{align}
  \partial_t \rho + \nabla \cdot (\rho \vec{V}) &= 0 \tag{18a} \\
  \partial_t(\rho \vec{V}) + \nabla \cdot \left[ \rho \vec{V V}
    + \left(p+B^{2}\right) \tens{I}-\vec{B B}\right] &= 0 \tag{18b} \\
  \partial_t \vec{B} + \nabla \cdot( \vec{V B} -\vec{B V}) &= 0 \tag{18c} \ .
\end{align}

\setcounter{equation}{18} The system is closed using the ideal gas

equation of state $p=\rho T$. The employed numerical scheme require
neither a Riemann solver nor a characteristic decomposition, and was
extended by means of a constrained transport description for the
magnetic field \citep[see, e.g.,][]{bals99, lond00}, which ensures
the solenoidality of the magnetic field. This method uses the
hyperbolic fluxes to compute the electric field components on a
staggered grid. These are then used to evolve the magnetic
induction, the components of which are also given on a (different)
staggered grid. The stability of the code and its capability to
resolve steep gradients without introducing spurious oscillations
have been demonstrated, e.g. by \citet{kiss06} and \citet{flaig09}.

In the employed code, the initial magnetic field is formulated by
means of a vector potential. The vector potential whose curl results in
the magnetic field of expression (\ref{eq:13}) can be expressed as

\begin{equation}
  \vec{A}_{{\rm in} \ r, \tilde{\varphi}, \tilde{z}}^{\prime}
  = B_{0, {\rm in}} \left[ 0,
    \ J_{1}\left(\frac{r}{\tilde{r_{0}}} \right),
    \ J_{0}\left(\frac{r}{\tilde{r_{0}}} \right) \right] \ .
\end{equation}

The vector potential corresponding to a radial magnetic field can be
expressed in the global spherical coordinates as

\begin{equation}
  \vec{A}_{1 R, \vartheta, \varphi} = \left[ 0, \
    -\frac{B_{0 \ {\rm out}} \Theta(\vartheta) \sin \vartheta}{R} \varphi, \
    0 \right] \ .
\end{equation}

In order to ensure continuity of the vector potential across the
surface of the MC, we represent the vector-potential of the
background magnetic field as

\begin{equation}
  \vec{A}_{\rm out} = \vec{A}_{1} + \nabla f
\end{equation}

with

\begin{equation}
  f = B_{0 {\rm in}} \left[
    r\left(1+\left(\frac{r_{0}}{r}\right)^{2}\right) +
    \tilde{r}_{0}J_{1}\left(\frac{r}{\tilde{r_{0}}}\right)\tilde{\varphi} +
    \tilde{r}_{0}J_{0}\left(\frac{r}{\tilde{r_{0}}}\right)\tilde{z}
  \right] \ .
\end{equation}

while the vector potential of the MC's magnetic field is given by

\begin{equation}
  \vec{A}_{\rm in} = \vec{A}_{\rm in}^{\prime} + \vec{A}_{\rm out} \ .
\end{equation}

\section{Details of the simulation setup}
In this section, we present details of the performed simulations.
The computational domain is a segment of a sphere bounded by
\mbox{$R \in [60,300]$}, \mbox{$\vartheta \in [0.1,0.9]\pi$}, and
\mbox{$\varphi \in [0,0.2]\pi$}. For the simulations we used 120 grid
cells in the radial and polar dimensions and 60 cells in the azimuthal
direction.\\
The initial velocity of the background flow is $V_{R}=20$,
corresponding to $410$~km/s. The center of the cylindrical MC is
initially located at $R_{0}=170$ and the MC's initial radial size
is $r_{0}=20$, with parameter $\tilde r_{0}=20$.
We conducted various simulation runs, which are summarized in
Table~\ref{table:1}.\\

\begin{table}
  \caption{Summary of conducted simulation runs, with references to the
    figures which display selected data from the respective runs.}
  \label{table:1}
  \centering
  \begin{tabular}{c c c c c}
    \hline\hline
    Run  & $V_{0}$ & $\omega$ & $B_{0out}$ & Figures\\
    \hline
    1  & 15 & 0    & $(200/R)^{2}$ & \ref{fig:1}, \ref{fig:6} \\
    2  & 20 & 0    & $(200/R)^{2}$ & \ref{fig:2}, \ref{fig:3}, \ref{fig:4} \\
    3  & 30 & 0    & $(200/R)^{2}$ & \ref{fig:5}, \ref{fig:6} \\
    4  & 20 & 0.05 & $(200/R)^{2}$ & \ref{fig:9} \\
    \hline
  \end{tabular}
\end{table}

In runs 1 to 3, we considered cases in which the MC's bulk velocity
$V_{0}$ is initialized to be lower than, equal to, or higher than
the ambient flow speed of 20. Furthermore, the (formerly vanishing)
MC's internal rotation around its axis of symmetry was varied, as
was the direction of the background B field (run 4). In all runs,
the mass density inside the MC was $\rho_{\rm in}=0.5$ and the
density outside the MC was $\rho_{\rm out}=(200/)R^2$. In order to
describe the magnetic field inside the MC, we set the value of
$B_{0,{\rm in}}=5$. The Lagrangian velocity of the MC's edge is
$V_{\rm S} = k r_{0}$ with $k=0.05$.

In order to ease a comparison with observations as well as with
analytical models, time profiles of density, magnetic field
magnitude, and flow velocity at selected fixed positions along the
MC's trajectory have been extracted from the simulation data and
presented in Figs.~\ref{fig:1}, \ref{fig:3}, \ref{fig:7}, and
\ref{fig:10}. These "virtual observers" thus capture information
which would be measured by a stationary spacecraft situated on the
MC's trajectory as the latter sweeps over it. All three ``virtual
observers'' have the same polar and azimuthal coordinates, viz.\
$\vartheta^{\prime} = 0.5\pi$ and $\varphi^{\prime} = 0.03\pi$, and
the respective distances of each of the three observers from the Sun
are $R_{1}^{\prime}=196$, $R_{2}^{\prime}=225$, and
$R_{3}^{\prime}=250$.

\section{Results and discussion}
\subsection{Non-rotating MCs}

In this section, we present and discuss the results of the numerical
simulations described in the previous section.

\begin{figure*}[t]
\vspace*{10mm}
\includegraphics[width=20cm]{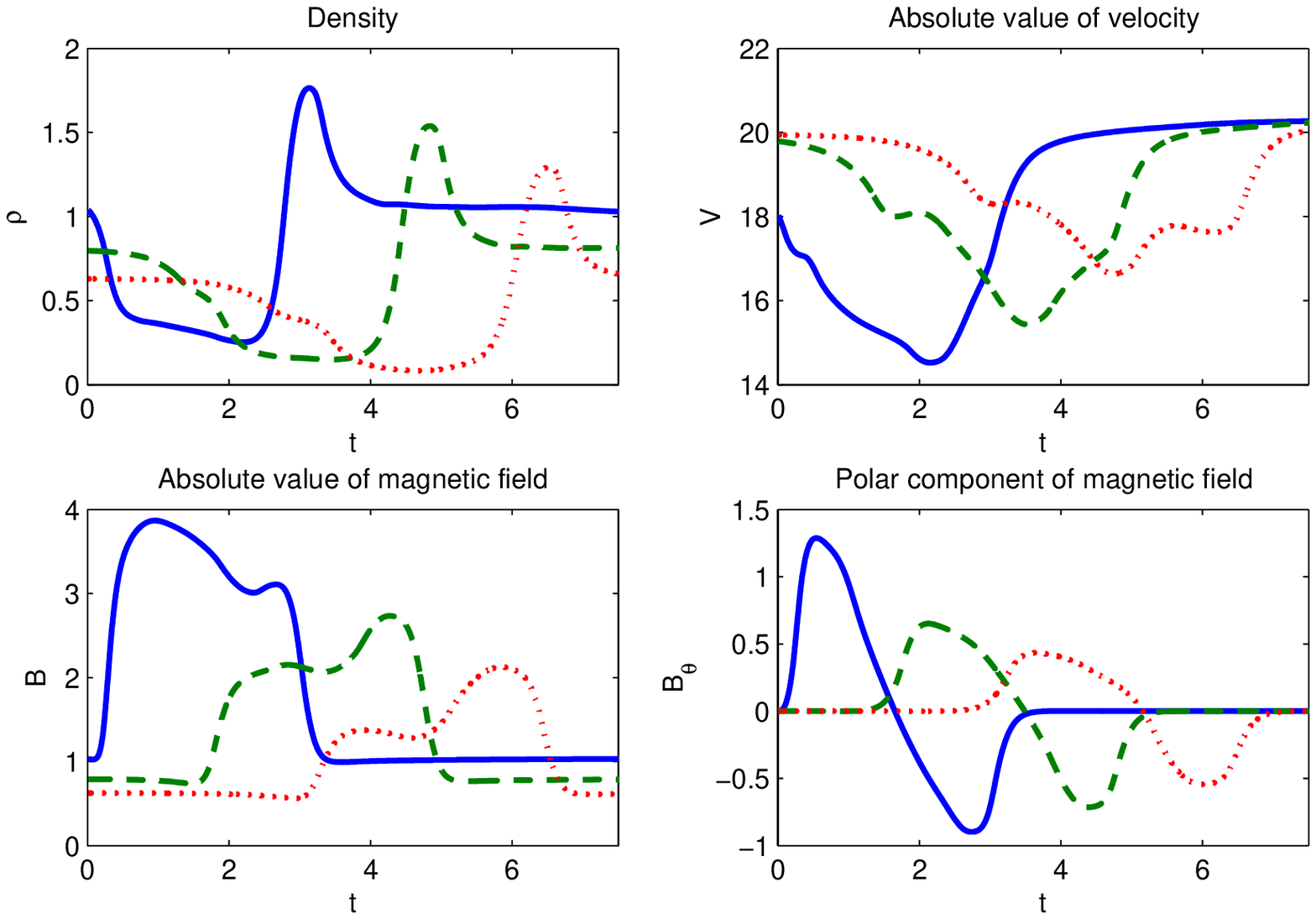}
\caption{ \label{fig:1}
  Plots of normalized density, velocity, total magnetic field, and polar
  component of the magnetic field. The solid, dashed, and dotted lines show
  the values recorded by virtual observers placed at $R=196$, $R=225$, and
  $R=250$, respectively. These plots correspond to run 1 (see
  Table~\ref{table:1}), for which $V_{0}=15$ and $\omega=0$.}
\end{figure*}

\begin{figure*}[t]
\vspace*{10mm}
\includegraphics[width=\textwidth]{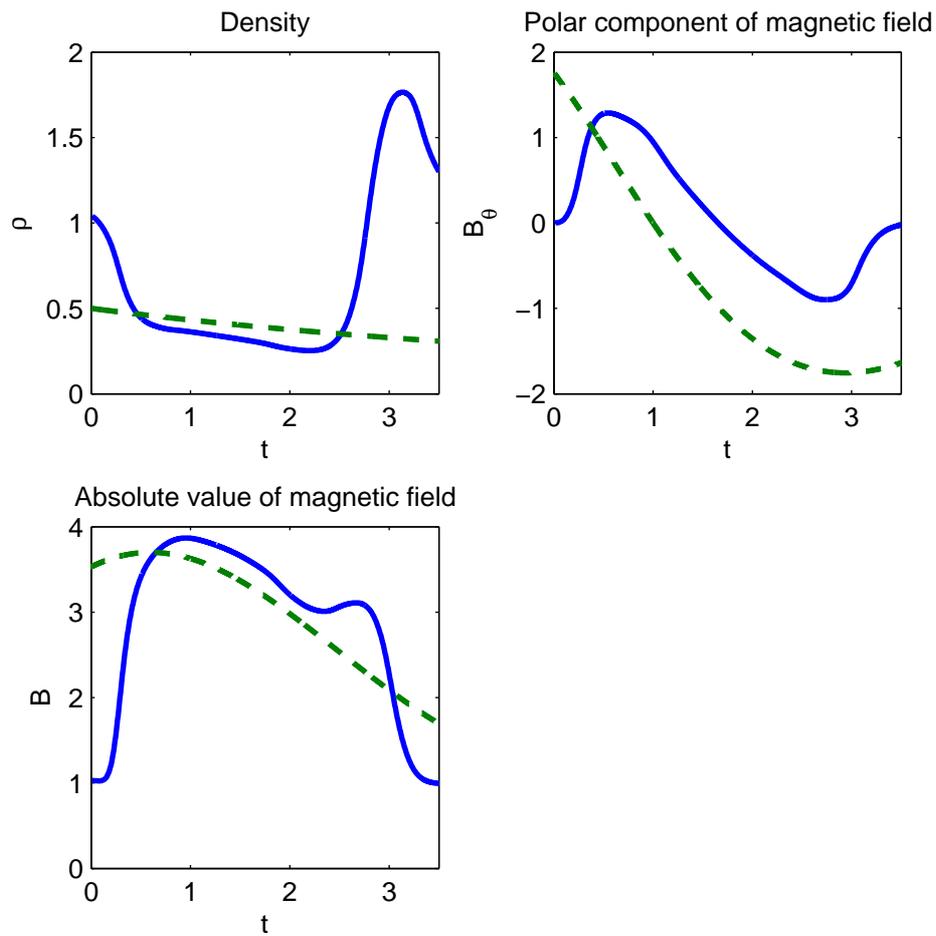}
 \caption{ \label{fig:2}
   Plots of density, polar component of magnetic field, and absolute value
   of the magnetic field. The solid lines correspond to the values detected by
   a virtual observer situated at $R=196$. The dashed line shows values
   obtained using the analytical solution. The initial bulk velocity of the
   MC is $V_{0}=15$, its initial radius is $r_{0}=20$.}
\end{figure*}

\begin{figure*}[t]
\vspace*{1mm}
\begin{center}
\includegraphics[width=20cm]{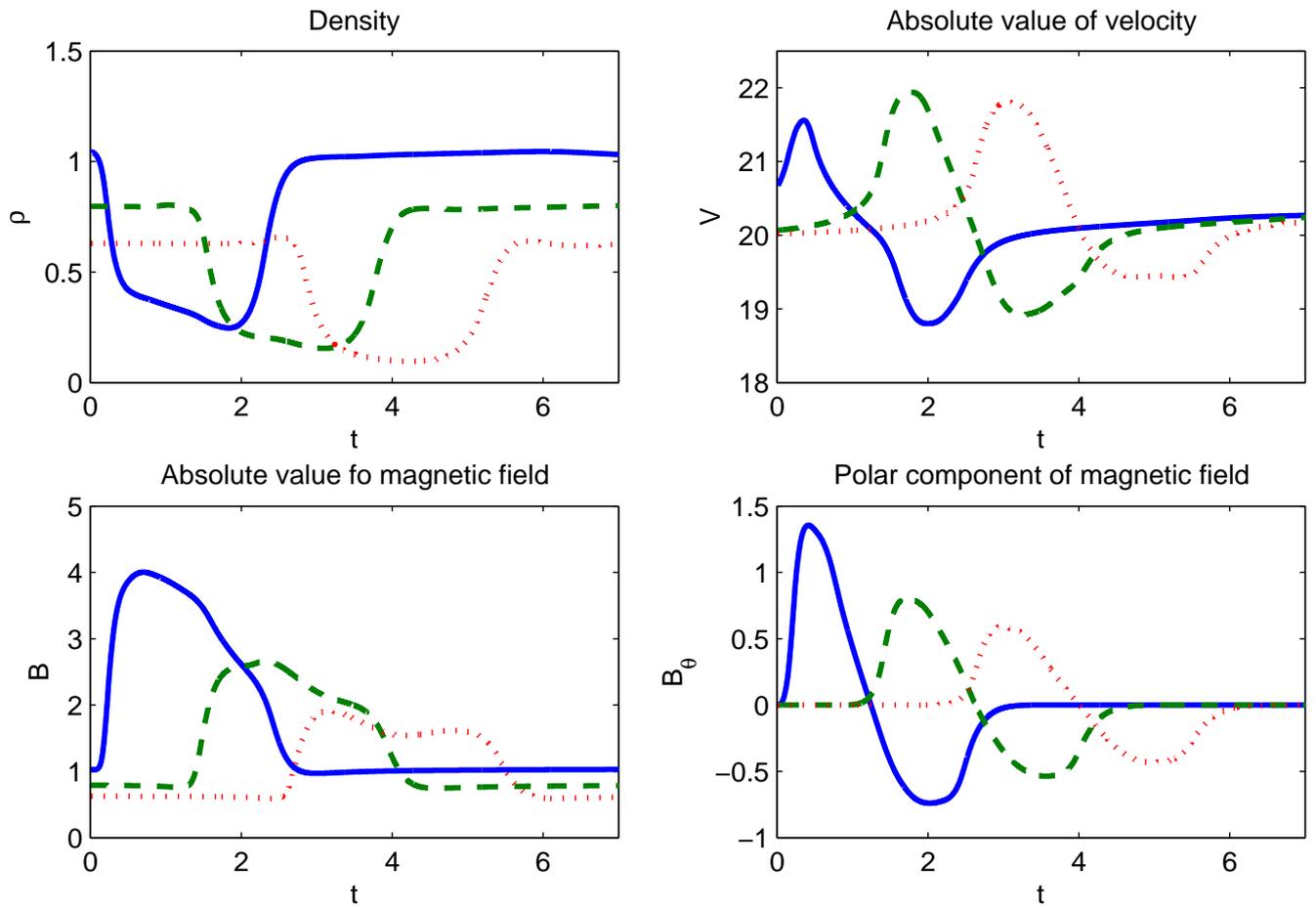}
\end{center}
\caption{ \label{fig:3}
  The same type of data as displayed as Fig.~\ref{fig:1}, here for run 2, when
  the MC's initial velocity is $V_{0}=20$ (i.e.\ equal to that of the
  ambient wind). We see that the structure maintains the MC's signatures
  during its evolution fairly well.}
\end{figure*}

\begin{figure*}[t]
\vspace*{10mm}
\includegraphics[width=\textwidth]{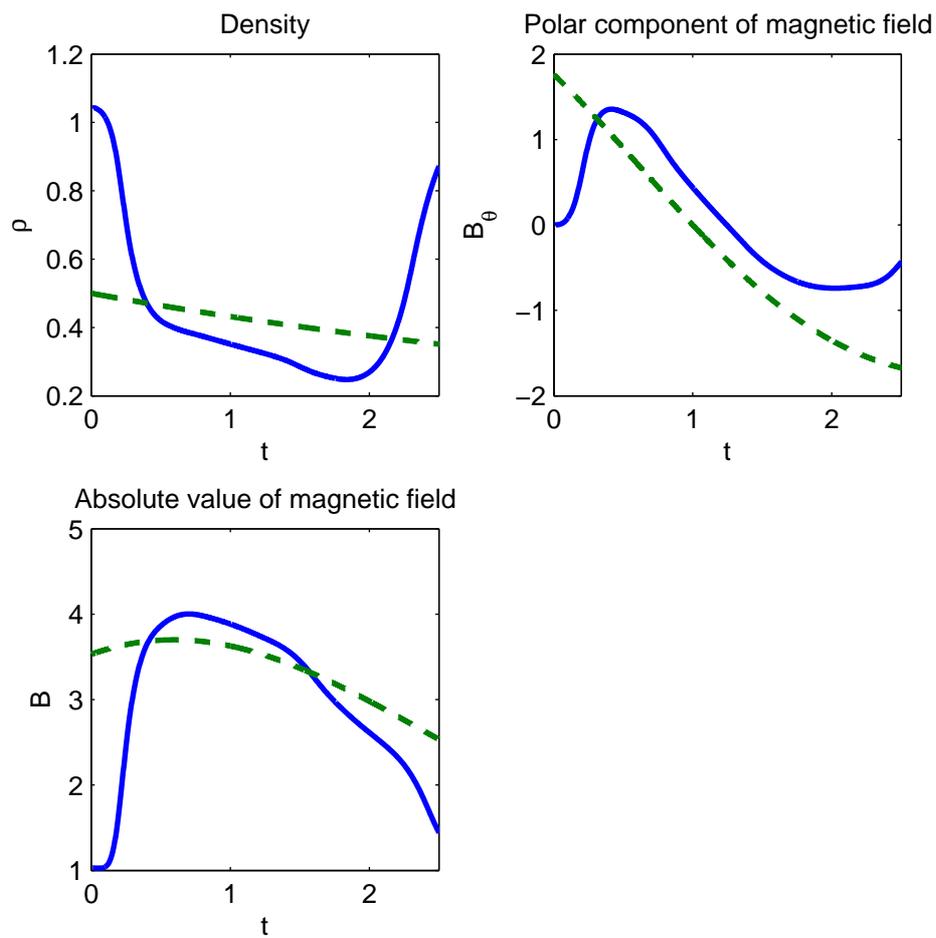}
 \caption{ \label{fig:4}
   The same plot as in Fig.~\ref{fig:2} but here for the case in which the
   MC's velocity is $V_{0}=20$ (i.e.\ equal to that of the background wind).}
\end{figure*}

\begin{figure*}[t]
\vspace*{1mm}
\begin{center}
\includegraphics[width=16cm]{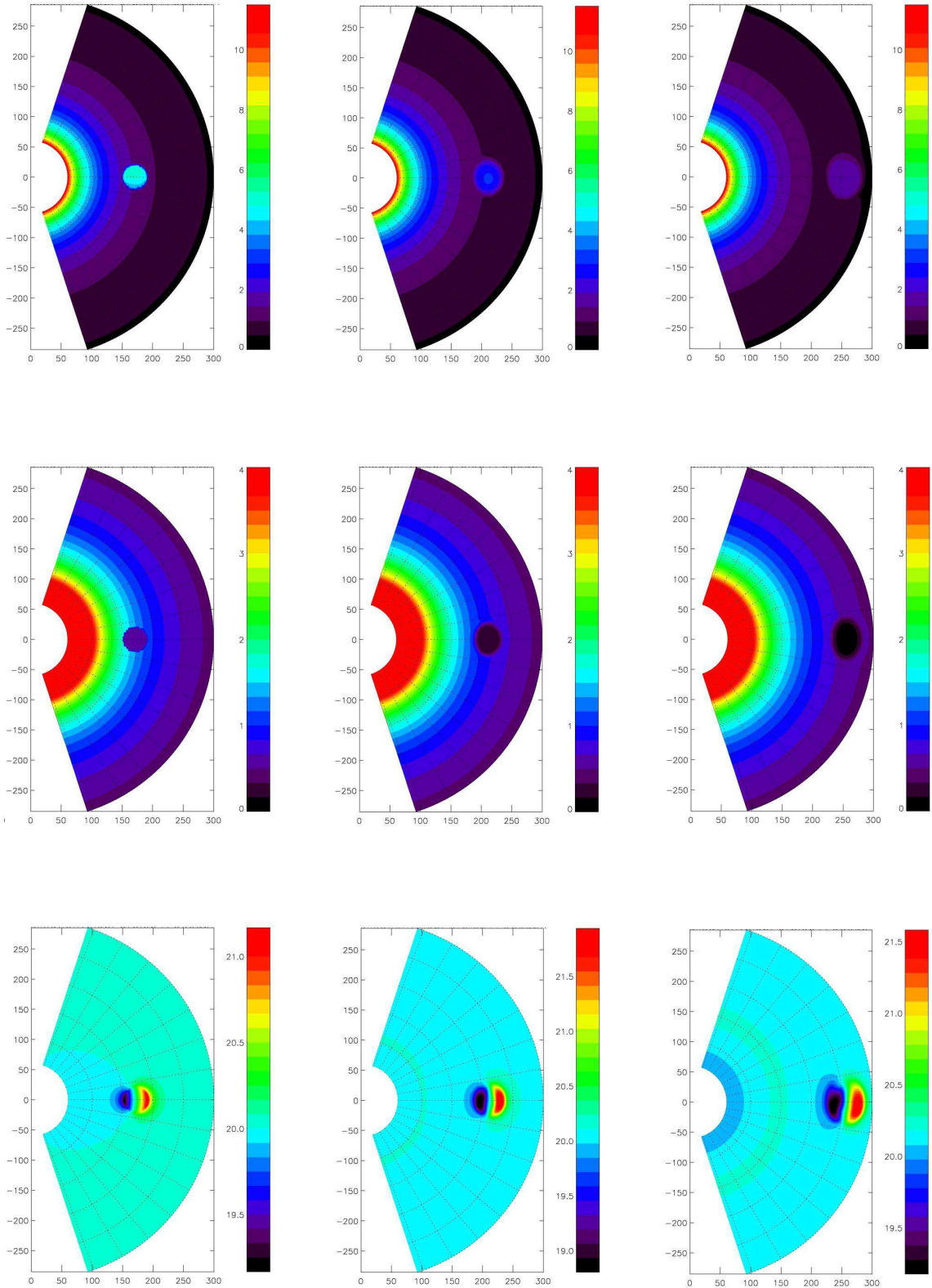}
\end{center}
\caption{ \label{fig:5}
  Cuts along the meridional plane $\varphi=0.03\pi$ for run 2
  (see Table~\ref{table:1}), displaying snapshots for the normalized
  values of the absolute magnetic field (top row), mass density (middle row),
  and absolute velocity (bottom row) at times $t=0$ (left column), $t=2$
  (middle column), and $t=4$ (right column). The MC's initial velocity
  is $V_{0}=20$ (i.e.\ equal to that of the background wind).
  The MC expands approximately symmetrically.}
\end{figure*}

\begin{figure*}[t]
\vspace*{1mm}
\begin{center}
\includegraphics[width=18cm]{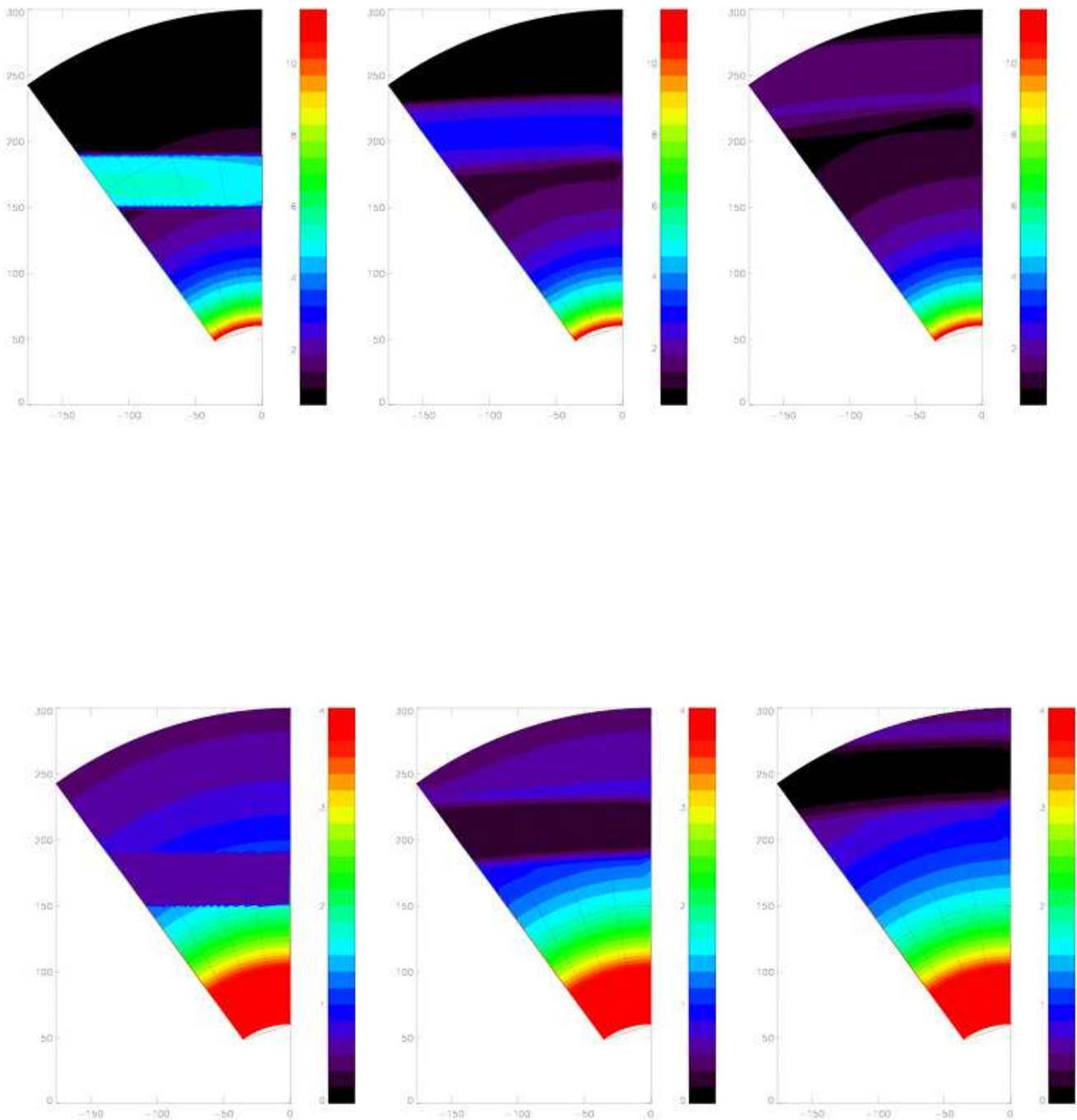}
\end{center}
\caption{ \label{fig:6}
  The same situation as depicted in Fig.~\ref{fig:3} (run 2,
  $V_{0}=20$, $\omega=0$), except a cut in the equatorial plane is shown. Top row corresponds to the global structure of the magnetic field and bottom row shows global structure of mass density.}
\end{figure*}

\begin{figure*}[t]
\vspace*{1mm}
\begin{center}
\includegraphics[width=18cm]{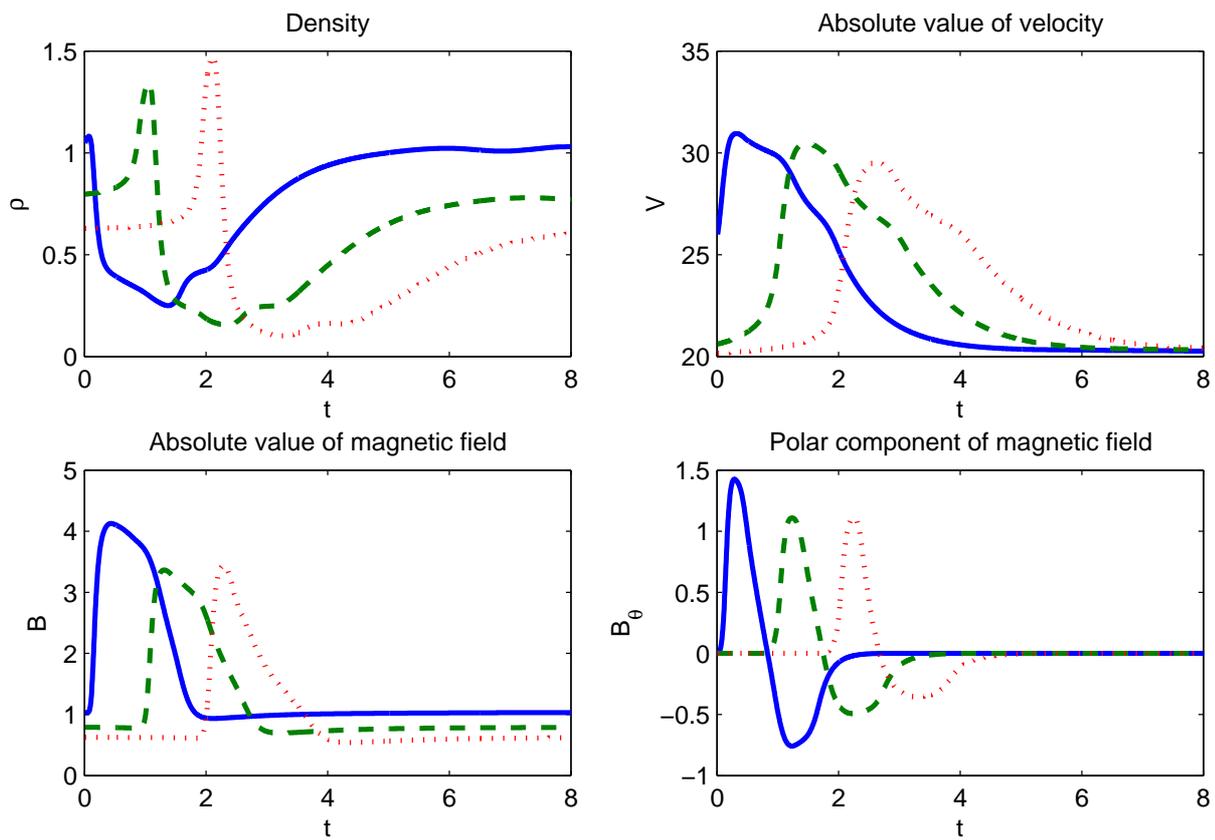}
\end{center}
\caption{ \label{fig:7}
  The same type of data as displayed in Fig.~\ref{fig:1}, here for run 3, when
  the MC's initial velocity is $V_{0}=30$ (i.e.\ higher than that of
  the background wind).}
\end{figure*}

\begin{figure*}[t]
\vspace*{10mm}
\includegraphics[width=\textwidth]{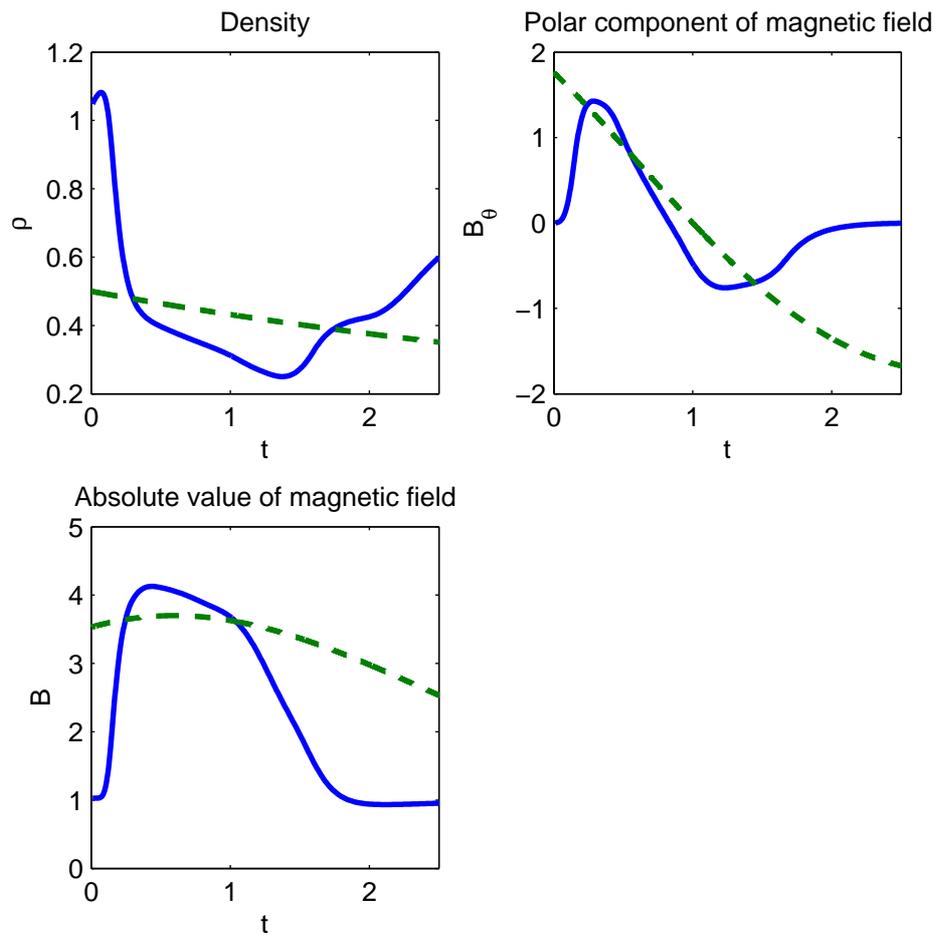}
 \caption{ \label{fig:8}
   The same plot as in Fig.~\ref{fig:2} but here for the case in which the
   MC's initial velocity is $V_{0}=30$ and its initial radius is $r_{0}=20$.}
\end{figure*}

\begin{figure*}[t]
\vspace*{1mm}
\begin{center}
\includegraphics[width=18cm]{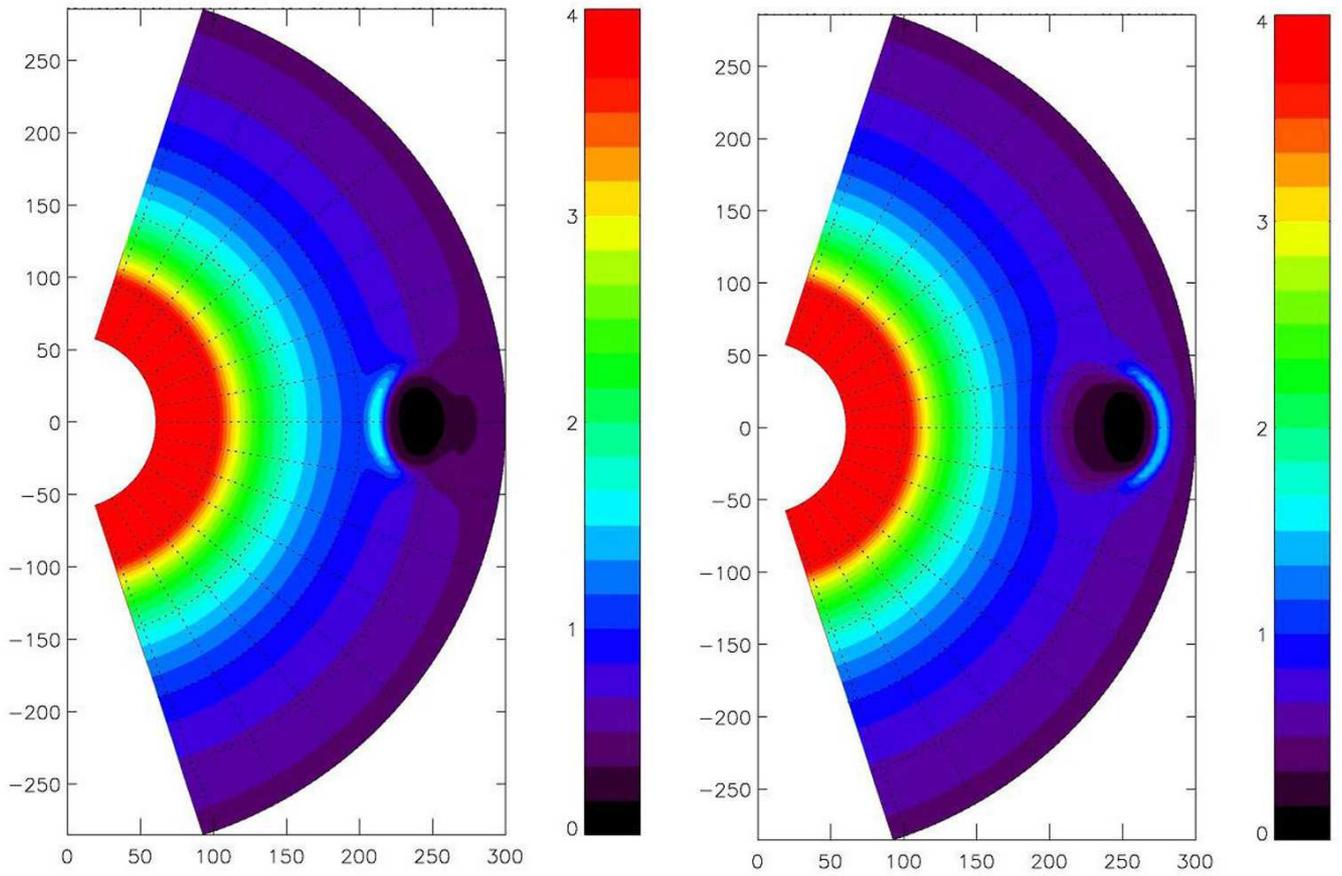}
\end{center}
\caption{ \label{fig:9}
  Density cuts along the meridional plane at time $t=2$ for run 1
  (left panel) and run 3 (right panel). In the left panel the MC initially
  moves with velocity $V_{0}=15$ (i.e.\ slower than the background wind).
  The right panel corresponds to the case in which the MC initially moves
  with velocity $V_{0}=30$ (i.e.\ faster than the background wind).
  We see that the slow MC is followed by a denser region while the fast MC
  is preceded by a region with a sharp density gradient.}
\end{figure*}

Figs.~\ref{fig:1}, \ref{fig:3}, and \ref{fig:7} show time profiles
of normalized density, velocity, total magnetic field, and polar
component of the magnetic field for the runs 1 to 3 as given in
Table~\ref{table:1} as recorded by the virtual observers. In
Figs.~\ref{fig:2}, \ref{fig:4}, and \ref{fig:8}, these numerical
results are compared with those obtained with the analytical model
for the region inside the MC, i.e.\ the physical variables are
plotted for the time interval from $t_{\rm in}$ when the observer
situated at $R=196$ enters the MC until $t_{fin}$, when this
observer leaves the MC again. $t_{in}$ and $t_{fin}$ are calculated
analytically. These comparisons reveal that the analytical model
describes the evolution of magnetic structures best for the case in
which the MC initially moves with the velocity of the ambient solar
wind. Fig.~\ref{fig:5} provides plots of the global structure of the
total magnetic field, mass density, and plasma velocity in a
meridional plane ($\varphi={\rm const}$) and on Fig.~\ref{fig:6} are
shown structures of magnetic field and mass density in the
equatorial plane. While these figures display the results for run 2,
Fig.~\ref{fig:9} shows the global structure of the number density in
the meridional plane for runs 1 and 3. From the results shown in
Figs.~\ref{fig:5} and \ref{fig:6} we can conclude that when the
velocity of the MC is close to the velocity of the background flow,
the MC expands smoothly, i.e.\ although it does not maintain an
exactly cylindrical shape (Fig.~\ref{fig:5}) and exhibits a slightly
changing axis curvature (Fig.~\ref{fig:6}), the plasma density
inside the MC decreases without developing strong gradients.
Fig.~\ref{fig:9} reveals that when the MC's velocity is less than
the velocity of the background flow, i.e.\ when the MC moves slower
than the background solar wind, a denser region appears behind the
MC after some time, while the MC which initially moves faster than
the ambient solar wind is preceded by a region with a sharp
(positive) density gradient. Observations do indeed show that MCs
moving faster than the ambient solar wind are preceded by density
enhancements \citep{kl82}. The simulation results  also show that
the MC exhibits stronger deviations from a cylindrical shape as
compared to the case shown in Figs.~\ref{fig:5} and \ref{fig:6}.
From a comparison of the upper left and bottom left panels of
Fig.~\ref{fig:1} (density and absolute value of the magnetic field),
one can see that after some time the observers detect an increase of
magnetic field and density at about the same time (see dashed and
dotted lines). Such structures that are characterized by sharp
gradient of density can also be found in observational data, see
\citep{lep97}. We could conclude that -- besides other factors such
as a possible overtaking of slow clouds by fast corotating streams
\citep{kl82} -- dynamical processes occurring during the interaction
of an MC with the ambient solar wind could play a crucial role for
the creation of sharp density gradients in the vicinity of an MC.
We also can see that the profile of the magnetic field strength
evolves asymmetrically, which could be explained by the fact that
the MC expands and, while an observer crosses this structure, the
magnetic field inside the MC does not stay constant, but decreases
in time. Note that very similar features are present in real
observations, as exemplified in \citep{lep97}. We can also see that
due to an increase of the MC's radius, the front parts of the MC are
observed to propagate with a higher velocity than its rear regions.
This feature is also observed in real data \citep[e.g.][]{nak08}.

From these results it is obvious that initially, in all three cases,
an observer would detect the following main signatures of a magnetic
cloud: 1)~a decrease of the plasma density, 2)~an increase of the
magnetic field strength, and 3)~a rotation of the magnetic field
inside the MC. We see that during the evolution not all magnetic
structures exhibit these main features. As a matter of fact, in the
cases when the velocity of the magnetic structure differs much from
the velocity of the background flow, the regions of lower density
and higher magnetic field evolve differently in time, see
Figs.~\ref{fig:1} and \ref{fig:7}. Only those structures that move with
a velocity close to that of the background flow maintain the
above-mentioned characteristic signatures.

\subsection{Rotating MCs}
\begin{figure*}[t]
\vspace*{1mm}
\begin{center}
\includegraphics[width=20cm]{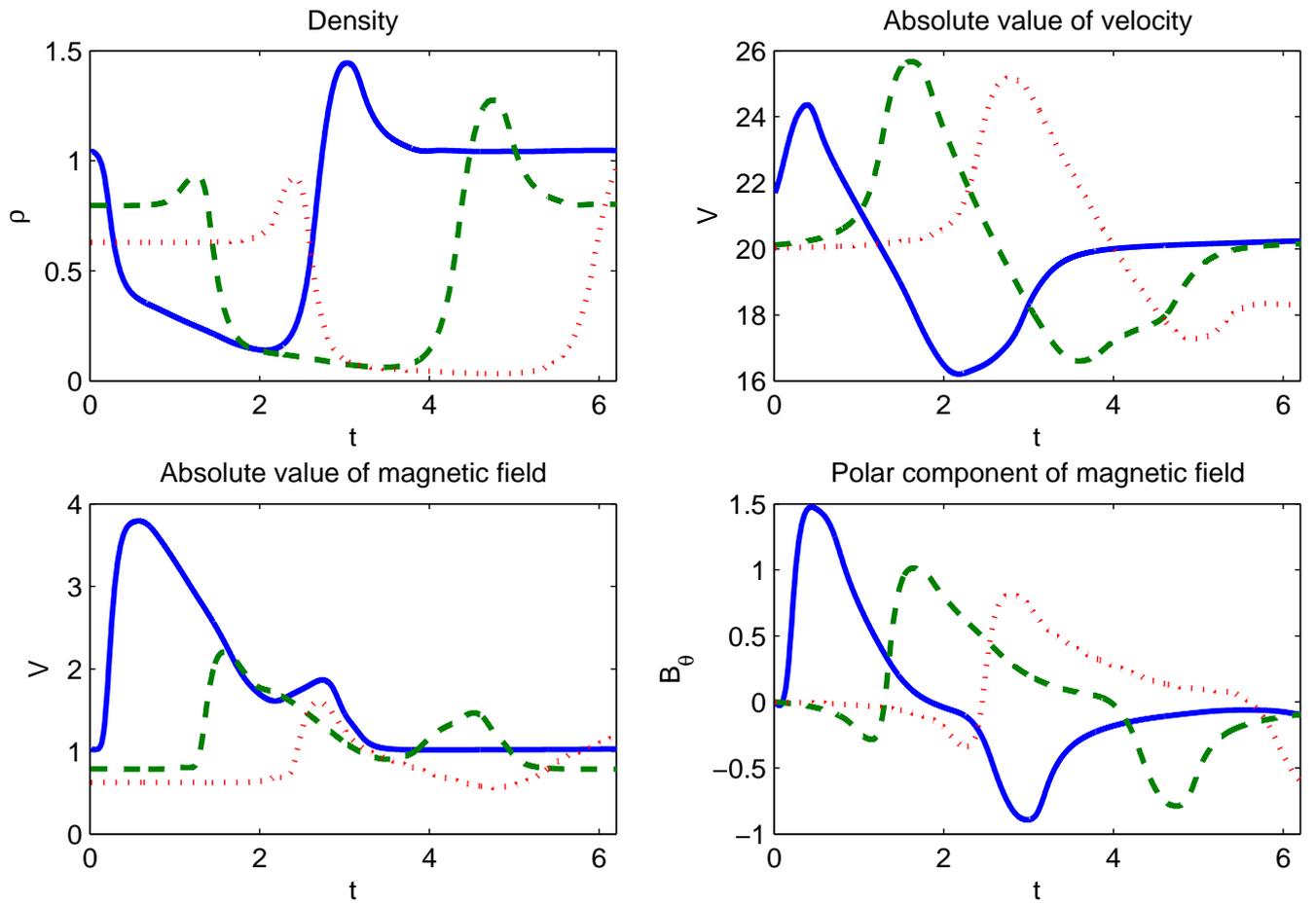}
\end{center}
\caption{ \label{fig:10}
  The same type of data display as Fig.~\ref{fig:1} for run 4
  ($V_{0}=20$, $\omega=0.05$).}
\end{figure*}

\begin{figure*}[t]
\vspace*{1mm}
\begin{center}
\includegraphics[width=20cm]{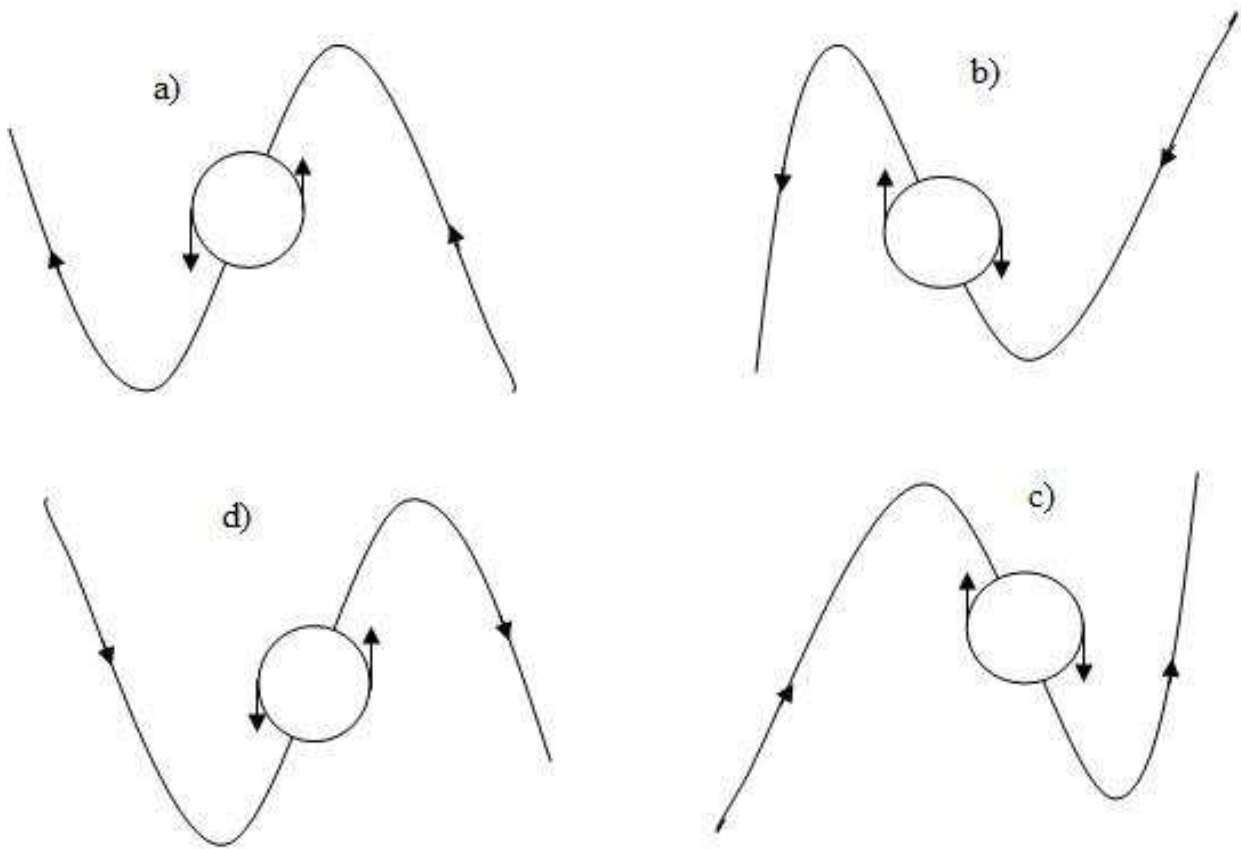}
\end{center}
\caption{ \label{fig:12}
  Illustration of the bending of a magnetic field line due to the rotation
  of the flux rope. Here it is assumed that solar wind flows from right to
  left. Panels a) and d) correspond to the north-south rotation of an MC,
  while panels b) and c) demonstrate the case of south-north rotation.
  Due to the MC's rotation field lines of the background magnetic field
  are curved.}
\end{figure*}

Fig.~\ref{fig:10} shows the MC properties for the case in which the
rotation of the MC around its axis of symmetry $\tilde{z}$,
  (see Fig.\ref{fig:11}) is taken into account.
In Fig.~\ref{fig:10} this is done for the case in which the
background magnetic field direction coincides with the direction of
the solar wind velocity and the MC rotates in the north-south
direction (see panel a in Fig.~\ref{fig:12}). In Fig.~\ref{fig:10},
peaks of density and magnetic field one be discerned both in front
of and at the rear of the MC. This can be understood as follows: A
rotating MC generates a centrifugal force, which induces centrifugal
motion of matter. One could thus expect an accumulation of mass at
the edges of the MC and, due to the frozen-in condition, the
magnetic field also moves and accumulates with the plasma. In the
panel showing the polar component $B_{\vartheta}$ of the magnetic
field, we can see that the rotation causes a bending of the ambient
magnetic field lines. For instance, in the case of a north-south
rotation and when the background magnetic field direction coincides
with the direction of solar wind velocity, $B_{\vartheta}$ becomes
negative in front of the MC, (See the bottom right panel in
Fig.~\ref{fig:10} and panel a in Fig.~\ref{fig:12}).

\section{Summary and conclusions}
In order to study the dynamics of MCs propagating in the solar wind,
we numerically implemented analytical expressions for the physical
variables characterizing an MC as initial conditions. These
expressions were derived for self-similarly evolving MCs and
introduced in a number of previous studies
\citep{bur88, lep90, far93, far95, nak08, dem09, dal10}.
\citet{van06, van09} used these functions to interpret particular
observations at a certain time.
We presented results describing the evolution of MCs for three
different cases, namely, when the bulk velocity of the MC is a)~less
than, b)~ equal to, and c)~faster than the velocity of the ambient
solar wind.
We found that the results of our numerical simulations to be in good
reasonable agreement with the observations.

It was demonstrated that the initially prescribed main signatures of
MCs, namely magnetic fields above ambient value, a mass density
lower than that in the ambient solar wind, and and a rotation of the
magnetic field, are best maintained in the case when the MC moves
with a velocity close to the velocity of the background flow. Due to
interaction with the ambient solar wind, initially slow CMEs are
accelerated, while fast CMEs slow down \citep{lyn03}. We could
expect that after sufficient time the velocity of solar eruption
will not differ much from the velocity of the ambient solar wind.
According to observations, approximate velocities of the MCs at 1~AU
are 400-450 km/s \citep{kl82}.


\citet{van09} compared observational data with
similar results obtained using an analytical approach and found them to
be in good agreement. In this work the authors employed the same
functions as we used in our present study as initial conditions.
Since our numerical results fit observations for different times, we
can further conclude that in a certain case, viz.\ when the bulk
velocity of the MC is close to the background flow velocity, the MC
evolves nearly self-similarly \citep[see also][]{dal10}.

We further studied cases in which an MC rotates around its axis. It
was argued that the centrifugal force leads to an accumulation of
matter and magnetic field at the edges of the MC. The rotation is
also able to cause a bending of the background magnetic field lines.
We see that when comparing the obtained results to observations, we
were able to demonstrate reasonable qualitative agreement. In
particular did we find cases in which the analytical models
advocated by a number of experts continue to be valid for the
description of MCs during their evolution, and such cases for which
this is not true.

While the numerical simulations led to interesting results which
compare more easily to observations than those from analytical
approaches, we have to admit that the presented model still contains
several idealizations. First of all, we introduced a much idealized
background flow, which was initialized by uniform radial flow
velocity, a radial magnetic field, and a spherically symmetric
distribution of the plasma density. However, the introduction of a
radial flow with a radial magnetic field has a logical base: a
number of numerical studies \citep{bart05, bart07, jacobs07, jens05,
jens09, asch08, ril08, amar10} show that when the solar wind beyond
the source surface reaches a steady state, it is characterized by
radial plasma flow and radial magnetic field without a latitudinal
component, though even if initially a more complicated magnetic
field were introduced. It might be advantageous to start the
simulations of the solar wind from specific initial conditions and
to implement a magnetic cloud only later, when the solar wind has
reached steady state conditions. However at this stage our numerical
facilities do not enable us to perform such
kind of simulations.\\
Second, we considered MCs as initially cylindrical and symmetric
structures. It would be interesting to also study the evolution of
other more complex magnetic structures, e.g.\ toroids, spheroids,
ellipsoids, as well as structures which remain connected to the Sun.
In our study, the MC is initially already located far from the Sun
and it has a certain bulk velocity. It would be reasonable to study
the self-consistent evolution of a solar eruption from the solar
surface until it reaches 1~AU.

Since the present study mainly intends to test an analytical model,
features of MCs were compared to observations only for the case of
non-rotating MCs. For the future, we plan to extend these
comparative studies also to rotating MCs. Also, in forthcoming work
we intend to study the dynamical evolution of MCs in different types
of background flow. Furthermore, we are also interested in the
numerical investigation of the interaction between several magnetic
structures: besides the MCs another class of solar ejecta, namely
``complex ejecta" was identified. While most of the magnetic clouds
were associated with a single CME, complex ejecta could have had
multiple sources. It was conjectured that some complex ejecta were
produced by the interaction of two or more halo CMEs \citep{bur02}.

\begin{acknowledgements}

We are grateful to Ralf Kissmann for providing his
numerical code and assistance. These results were obtained in the
framework of the project by European Commission through the SOLAIRE
Network (MTRN-CT-2006-035484). Financial support of Forschergruppe
1048 (project FI 706/8-1) funded by the Deutsche Forschungsgemeinschaft
(DFG), GOA/2009-009 (K.~U.~Leuven), G.0304.07 (FWO-Vlaanderen) and
C~90347 (ESA Prodex 9) is acknowledged.

\end{acknowledgements}

\bibliographystyle{aa}
\bibliography{references_magcloud}

\newpage

\newpage


\newpage


\newpage

\newpage





\newpage
\newpage

\end{document}